\documentclass
[aps,preprint,tightenlines,superscriptaddress,bibnotes,balancelastpage,secnumarabic,titlepage,twocolumn,10pt]{revtex4}%
\usepackage{amsfonts}
\usepackage{amsmath}
\usepackage{amssymb}
\usepackage{graphicx}%
\providecommand{\U}[1]{\protect\rule{.1in}{.1in}}

\begin{document}
\preprint{ANL-HEP-PR-11-69 and UMTG-25}
\title[RG -- A \& C]{RG flows, cycles, and c-theorem folklore}
\author{Thomas L. Curtright}
\affiliation{Department of Physics, University of Miami, Coral Gables, FL 33124-8046, USA}
\author{Xiang Jin}
\affiliation{Department of Physics, University of Miami, Coral Gables, FL 33124-8046, USA}
\author{Cosmas K. Zachos}
\affiliation{High Energy Physics Division, Argonne National Laboratory, Argonne, IL
60439-4815, USA\medskip\medskip}

\begin{abstract}
Monotonic renormalization group flows of the \textquotedblleft
c\textquotedblright\ and \textquotedblleft a\textquotedblright\ functions are
often cited as reasons why cyclic or chaotic coupling trajectories cannot
occur. \ It is argued here, based on simple examples, that this is not
necessarily true. \ Simultaneous monotonic and cyclic flows can be compatible
if the flow-function is multi-valued in the couplings.

\end{abstract}
\volumeyear{year}
\volumenumber{number}
\issuenumber{number}
\eid{identifier}
\startpage{1}
\endpage{ }
\maketitle

Exact general results for renormalization group (RG) flows are important as
they may provide physical insight for strongly coupled systems. \ The
c-theorem for 2D systems \cite{Zamolodchikov} and the a-theorem for 4D systems
\cite{Cardy, Anselmi et al.} are two such results that have been established
for very broad classes of models \cite{Schwimmer}.

The c-theorem shows the existence of a monotonically decreasing function of
the length scale, $c\left(  L\right)  $, which interpolates between 2D
Virasoro central charges of theories at conformal fixed points, and thereby
provides an intuitively correct count of system degrees of freedom --- fewer
in the infrared than in the ultraviolet. \ The a-theorem establishes similar
monotonic flow for the induced coefficient of the Euler density, $a\left(
L\right)  $, for a 4D theory in a curved spacetime background. \ 

It is a common conclusion --- a \textquotedblleft folk
theorem\textquotedblright\ --- based on these monotonically evolving
\textquotedblleft observables\textquotedblright\ that the underlying couplings
can \emph{not} have RG trajectories which are limit cycles or undergo more
exotic (e.g. chaotic) oscillations (e.g. see 2nd bullet item under \S 6 in
\cite{Cardy2010}). \ The point of this note is to explain and illustrate with
just one coupling, as simply as possible, why this conclusion is unwarranted.
\ (Somewhat similar criticism of the monotonic folklore has been proffered in
other contexts, involving degenerate Morse function counterexamples for models
with\ vorticity in the flow of several couplings \cite{Niemi}.)

In principle, we believe cyclic or perhaps even chaotic coupling trajectories
are \emph{not} ruled out by either the $c$- or $a$-theorems, nor are they
necessarily excluded by other monotonic \textquotedblleft potential
flow-functions.\textquotedblright\ \ To illustrate our reasoning, we begin
with a very simple example based on a mechanical analogy. \ While this example
does indeed exhibit both monotonic flow and a cycling trajectory, it has the
peculiar feature --- insofar as intuitively counting degrees of freedom is
concerned --- that the monotonic flow is unbounded both above and below.
\ Nevertheless, we recall there is a field theory model that produces just
such behavior \cite{LeClair}. \ We then exhibit another example where the
monotonic flow is bounded below and the coupling trajectory is not only cyclic
but, in fact, chaotic.

The essential ideas, expressed for a single coupling $x\left(  t\right)  $,
where $t=\ln L$, are given by general statements for a locally gradient RG
flow,%
\begin{align}
\frac{dx\left(  t\right)  }{dt}  &  =\beta\left(  x\left(  t\right)  \right)
=-\frac{dC\left(  x\left(  t\right)  \right)  }{dx\left(  t\right)  },\\
\frac{dC\left(  x\left(  t\right)  \right)  }{dt}  &  =\frac{dx}{dt}\frac
{dC}{dx}=\beta~\frac{dC}{dx}=-\left(  \frac{dC}{dx}\right)  ^{2},
\end{align}
and by a specific example of a flow-function, namely,%
\begin{equation}
C_{0}\left(  x\right)  =-\frac{\pi}{4}-\frac{1}{2}\arcsin\left(  x\right)
-\frac{1}{2}~x\sqrt{1-x^{2}}.
\end{equation}
The corresponding $\beta$ function is
\begin{equation}
\beta_{0}\left(  x\right)  =-\frac{d}{dx}~C_{0}\left(  x\right)
=\sqrt{1-x^{2}}.
\end{equation}
The RG flow is given by%
\begin{equation}
\frac{dx}{dt}=\sqrt{1-x^{2}},
\end{equation}
which is easily recognized as a \textquotedblleft
right-moving\textquotedblright\ simple harmonic oscillator (SHO) started from
rest at $x=-1$. \ This of course has a turning point, $x=+1$, reached in
finite $\Delta t$, at which point the only way to continue the evolution is to
change branches of the square root, $\sqrt{1-x^{2}}\rightarrow-\sqrt{1-x^{2}}%
$, to produce a \textquotedblleft left-moving\textquotedblright\ SHO. \ When
this procedure is repeated as turning points are encountered, the cyclic
evolution emerges.

In addition, when the first turning point is encountered $C$ switches to a
second branch, given by
\begin{equation}
C_{1}\left(  x\right)  =-\frac{3\pi}{4}+\frac{1}{2}\arcsin\left(  x\right)
+\frac{1}{2}~x\sqrt{1-x^{2}}.
\end{equation}
This gives the expected switch between branches for the $\beta$ function,%
\begin{equation}
\frac{dx}{dt}=-\frac{d}{dx}~C_{1}\left(  x\right)  =-\sqrt{1-x^{2}}\text{.}%
\end{equation}
More importantly, this $C$ function continues to decrease monotonically
\emph{as a function of} $t$ after switching branches. \ 

This is easily understood for this simple example just because the
monotonically changing $C$ is nothing but the negative of the definite
integral of \textquotedblleft the oscillator's kinetic
energy\textquotedblright\ $T=\left(  dx/dt\right)  ^{2}$,
\begin{equation}
C=-\int\beta dx=-\int_{x\left(  0\right)  =-1}^{x\left(  t\right)  }\frac
{dx}{dt}dx=-\int_{0}^{t}Tdt, \label{AA}%
\end{equation}
where the integral is taken along the actual trajectory of the oscillator ---
a path that conserves total \textquotedblleft energy,\textquotedblright\ cf.
RG invariants. \ (That is to say, $C$ is just the \emph{reduced}\ or
\emph{abbreviated\ action} of Euler, Maupertuis, and Lagrange, or perhaps more
consistently with the notation, it is the \emph{characteristic function} of Hamilton.)

In fact, to obtain the correct evolution for the continuous flow in question,
it is absolutely necessary not only to switch between the two branches for
$\beta\left(  x\right)  =\pm\sqrt{1-x^{2}}$, but also to switch among an
\emph{infinite} set of branches for the $C$-function, as successive turning
points are encountered. \ Thus, as an analytic function, $C$ involves a
nontrivial Riemann sheet structure \cite{Meurice}. \ With initial flow to the
right, $\left.  dx/dt\right\vert _{t=0}>0$, after $N$ encounters with turning
points, the evolution is given by%
\begin{align}
\frac{dx}{dt}  &  =\left(  -\right)  ^{N}\sqrt{1-x^{2}}=-\frac{d}{dx}%
C_{N}\left(  x\right)  ,\\
C_{N}\left(  x\right)   &  =-\frac{\pi}{4}\left(  1+2N\right) \nonumber\\
&  -\left(  -1\right)  ^{N}\left(  \frac{1}{2}\arcsin\left(  x\right)
+\frac{1}{2}~x\sqrt{1-x^{2}}\right)  ,
\end{align}
where $\arcsin$ is the principal branch of the inverse sine function. \ We
plot a few branches of $C$ in Figure 1.%
\begin{figure}[ptb]%
\centering
\includegraphics[
height=2.1871in,
width=3.2958in
]%
{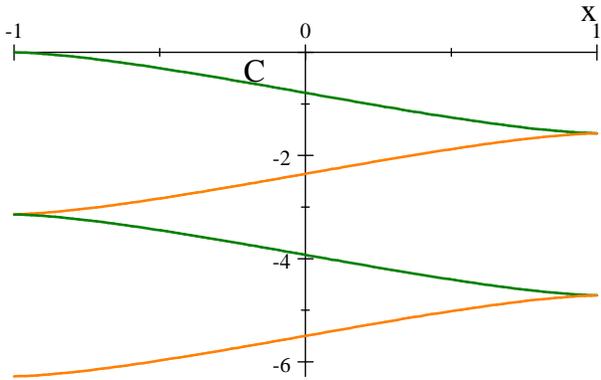}%
\caption{Four branches of the SHO $C\left(  x\right)  $ function.}%
\end{figure}
More directly, as a function of $t$,
\begin{equation}
C\left(  t\right)  =-\frac{1}{2}\left(  t-\cos t\sin t\right)  ,
\end{equation}
which is indeed monotonic in $t$, as shown in Figure 2.
\begin{figure}[ptb]%
\centering
\includegraphics[
height=2.1871in,
width=3.2958in
]%
{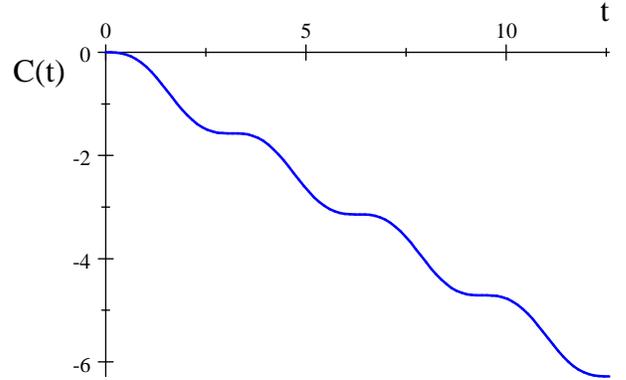}%
\caption{Monotonic flow for the SHO $C\left(  t\right)  $.}%
\end{figure}

The SHO example of simultaneous monotonic and cyclic flows, while certainly
familiar, is perhaps disconcerting, not just because of the multi-valuedness
of $C\left(  x\right)  $, but also because $C\left(  t\right)  $ is unbounded
both above and \emph{below}. \ However, this same cyclic flow may also be
observed by selecting different coordinates for the coupling, without changing
the physics of the system. \ Indeed, the \textquotedblleft Russian doll
superconductivity model\textquotedblright\ of Leclair et al.
\cite{LeClair,Braaten} provides a single flowing coupling $u$ that illustrates
what we have in mind. \ For that model the RG $\beta$ and corresponding $C$
function are given by innocuous polynomials,
\begin{equation}
\frac{du}{dt}=\frac{1}{2}\left(  1+u^{2}\right)  \ ,\ \ \ \mathcal{C}%
=-\frac{1}{2}~u\left(  1+\frac{1}{3}u^{2}\right)  .
\end{equation}
Despite this uncomplicated local behavior, the global trajectories go through
infinite excursions in the course of their cyclic evolution:%
\begin{equation}
u\left(  t\right)  =\tan\left(  \frac{1}{2}~t+\arctan u\left(  0\right)
\right)  .
\end{equation}
Thus it is difficult to keep track of the monotonicity of $\mathcal{C}$, if
any, as it executes an infinite jump during the course of each cycle.

The system is perhaps easier to grasp upon being expressed in terms of a
\textquotedblleft dual\textquotedblright\ coupling, $x$,
\begin{equation}
u=\pm\sqrt{\frac{1+x}{1-x}}\ ,\ \ \ \frac{dx}{dt}=\pm\sqrt{1-x^{2}}.
\end{equation}
That is to say, \emph{the RG flow of the model is equivalent to the SHO} as
described earlier. \ Note the cyclic switching between the branches of
$u\left(  x\right)  $ corresponding to right-moving (green) and left-moving
(orange) SHO motion, including an infinite jump upon reaching $x=1$, as shown
in Figure 3.%
\begin{figure}[ptb]%
\centering
\includegraphics[
height=2.1871in,
width=3.2958in
]%
{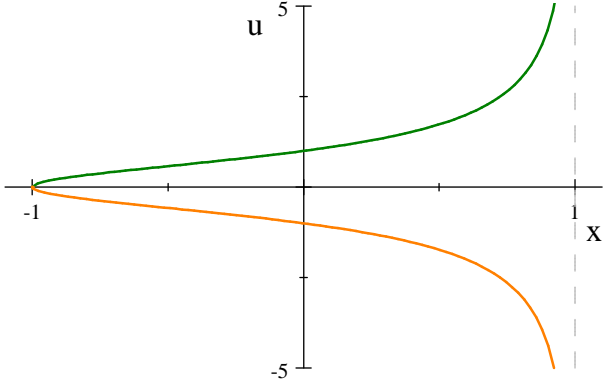}%
\caption{The Russian doll -- SHO RG duality.}%
\end{figure}

Similar analysis can be carried out for theories with several coupling
constants. \ (For models with limit cycles in $4-\varepsilon$ dimensions, see
\cite{Grinstein,Nakayama}.) \ We leave the study of these for another venue.

To complete this brief discussion, we consider a model with a cyclic but
chaotic trajectory which also exhibits a monotonic flow-function. \ Again, a
solvable example involving a single coupling is sufficient to make the point.

Perhaps the simplest system with chaotic RG evolution is the Ising model with
imaginary magnetic field, described by the special case of the logistic map
with parameter $4$ \cite{Dolan,CZ}. \ The \emph{exact} trajectory and $\beta$
function are given by%
\begin{align}
x\left(  t\right)   &  =\left(  \sin\left(  2^{-t}\arcsin\sqrt{x}\right)
\right)  ^{2}\ ,\\
\frac{dx\left(  t\right)  }{dt}  &  =-\left(  \ln4\right)  \sqrt{x\left(
t\right)  \left(  1-x\left(  t\right)  \right)  }\arcsin\sqrt{x\left(
t\right)  }\ ,
\end{align}
where the $\arcsin$ function in this last expression switches branches upon
encountering turning points. \ Similarly, the corresponding $C$ function,
considered as a function of $x\left(  t\right)  $, also changes branches at
turning points.

The direction of the flow in $t$ is such that the origin is an attractive
fixed point in the infrared, so $x\rightarrow0$ as $L$ \& $t=\ln
L\rightarrow+\infty$. \ On the other hand, $x$ becomes chaotic, exhibiting
cycles of arbitrary length, as $L\rightarrow0$ and $t\rightarrow-\infty$.
\ That is to say, for any initial $x\in\left(  0,1\right]  $ the flow for
$t>0$ is monotonically toward the fixed point at $x=0$, while for $t<0$ the
flow is toward a turning point at $x=1$, where $dx/dt$ reverses and the flow
is toward a second turning point at $x=0$ --- the zero of $\beta$ at $x=0$ is
a fixed point only for the first branch of $\beta$. \ As the evolution
continues into the UV, with $t<0$, the trajectory oscillates between the pair
of turning points, $x=0$ and $x=1$, with increasing average \textquotedblleft
speed.\textquotedblright

There are an infinite number of branches for both $\beta\left(  x\right)  $
and $C\left(  x\right)  $ in this case. \ Those branches are given by%
\begin{gather}
\beta_{N}\left(  x\right)  =-\left(  \ln4\right)  \sqrt{x\left(  1-x\right)
}\left\{  \left(  -\right)  ^{N}\left\lfloor \tfrac{1+N}{2}\right\rfloor
\pi+\arcsin\sqrt{x}\right\}  ,\nonumber\\
C_{N}\left(  x\right)  =\frac{1}{8}\left(  \ln4\right)  \left\{  {}\right.
4x^{2}\left(  x-1\right)  ^{2}\\
+\left(  \sqrt{x\left(  1-x\right)  }\left(  1-2x\right)  -\left(  -\right)
^{N}\left\lfloor \tfrac{1+N}{2}\right\rfloor \pi-\arcsin\sqrt{x}\right)
^{2}\left.  {}\right\}  .\nonumber
\end{gather}
Here $\arcsin$ is understood to be the principal branch, $\left\lfloor
\cdots\right\rfloor $ is the floor function, and $N$ counts the number of
encounters with the trajectory turning points at $x=1$ and $x=0$. \ The first
three branches of $C\left(  x\right)  $ are shown in Figure 4. \
\begin{figure}[ptb]%
\centering
\includegraphics[
height=2.1871in,
width=3.2958in
]%
{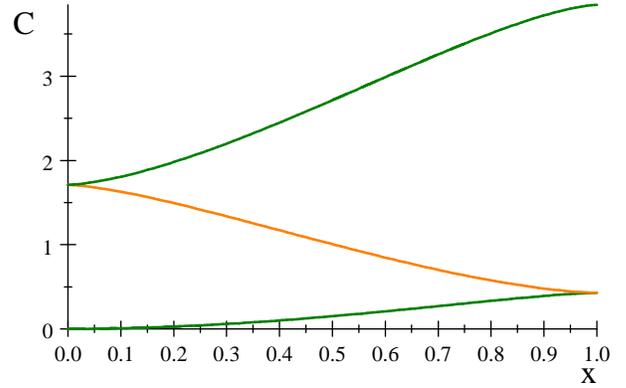}%
\caption{Three branches of the logistic $C\left(  x\right)  $ function.}%
\end{figure}
As $t\rightarrow\infty$, the flow is toward the origin, with $x\left(
+\infty\right)  =0$ and $C\left(  +\infty\right)  =0$, while as $t\rightarrow
-\infty$, $C\rightarrow+\infty$. \ This is more clearly seen by plotting%
\begin{equation}
C\left(  t\right)  =-\int_{0}^{x\left(  t\right)  }\beta\left(  x\right)
dx=\int_{t}^{\infty}\left(  \beta\left(  x\left(  t\right)  \right)  \right)
^{2}dt\ ,
\end{equation}
for $0<\left.  x\left(  t\right)  \right\vert _{t=0}<1$. \ The flow of $C$ is
monotonic in $t$ and bounded below, $C\geq0$. \ This is shown in Figure 5 for
$\left.  x\left(  t\right)  \right\vert _{t=0}=1/2$.%
\begin{figure}[ptb]%
\centering
\includegraphics[
height=2.1871in,
width=3.2958in
]%
{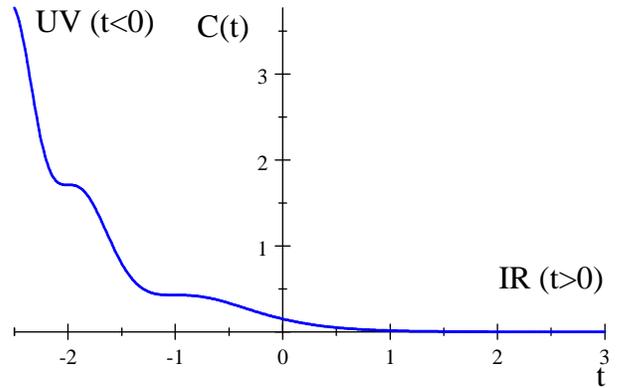}%
\caption{Monotonic flow for the logistic $C\left(  t\right)  $.}%
\end{figure}

A full discussion of Lagrangian models that realize this second example will
have to be given elsewhere. \ Suffice it to say here that chaotic RG
trajectories have indeed appeared in spin-glass systems \cite{McKay,Bray}.
\ The point we wish to emphasize is that such behavior is not necessarily
inconsistent with $c$- and $a$-theorems.

In conclusion, we have argued against the folklore\ that cyclic RG
trajectories are always incompatible with a monotonic potential flow-function
producing a gradient flow. \ We have shown by examples that monotonic
evolution of $C\left(  t\right)  $ can be consistent with cyclic coupling
trajectories given by gradient flows, when the flow-function $C$ is
multi-valued in the couplings.

\textbf{Acknowledgments} \ \textit{We thank D Z Freedman, Ian Low, and Y
Meurice for helpful discussions and critical remarks. \ This work was
supported in part by NSF Award 0855386, and in part by the U.S. Department of
Energy, Division of High Energy Physics, under contract DE-AC02-06CH11357.}

\end{document}